# Fabrication of high-entropy $RE$Ba$_2$Cu$_3$O$_{7-\delta}$ thin films by pulsed laser deposition

Aichi Yamashita[1*], Kazuki Hashimoto[1], Yusuke Nakanishi[1], Toshihiko Maeda[2], Yoshikazu Mizuguchi[1]

[1]Department of Physics, Tokyo Metropolitan University, Hachioji 192-0397, Japan
[2]School of Environmental Science and Engineering, Kochi University of Technology, Kami 782-8502, Japan

*Corresponding author: Aichi Yamashita (aichi@tmu.ac.jp)

Epitaxial thin films of $RE$Ba$_2$Cu$_3$O$_{7-\delta}$ ($RE$123, $RE$: rare earth) having a high-entropy (HE) $RE$ site were successfully fabricated on a SrTiO$_3$ substrate by the pulsed laser deposition method. One to five $RE$ elements are solved at the $RE$ site, which results in an increase in configurational entropy of mixing ($\Delta S_{mix}$). Through the measurements of critical current density ($J_c$) by a magnetization and the Bean's model analysis, we found that the $J_c$ of the HE films exceeds an order of 1.0 MA/cm$^2$ under conditions of $T < 20$ K and $\mu_0 H < 7$ T. Since the HE effects have a potential to an improvement of irradiation tolerance, the present results encourage further development of HE $RE$123 superconducting materials for a practical use in the environment with high magnetic fields and irradiation, for example in fusion reactors.

Since the discovery of high-transition temperature (high-$T_c$) cuprate superconductor[1], various kinds of cuprate superconductors have been reported[2-5]. Among them, $RE$Ba$_2$Cu$_3$O$_{7-\delta}$ ($RE$123, $RE$: rare earth) is a promising material for high-field superconductivity application because of its high-$T_c$ and high critical current density (high-$J_c$) under magnetic fields in a thin film form[6]. These characteristics enable us to practically use $RE$123 superconductors under high temperatures and high magnetic fields, which are not capable by low-$T_c$ superconductors such as NbTi and Nb$_3$Sn. High-$T_c$ cuprate superconductors are desired as a promising candidate for a magnet material in a next-generation energy reactor[7], because of their high critical magnetic field more than 20 T could realize the compact nuclear fusion design and operating at relatively high temperature such as $T = 20$ K, which are not possible by low-$T_c$ superconductors. In a nuclear fusion reaction, neutrons are generated and irradiate surrounding superconducting magnets. In some practical superconductors such as Nb$_3$Sn and $RE$Ba$_2$Cu$_3$O$_{7-\delta}$, it has been revealed that the superconducting characteristics obviously deteriorate by neutron and high-energy particle irradiation[8,9]. Therefore, development of superconducting material sufficiently resistant to the environment under high-energy particle irradiation is one of the most important issues for the safe operation of the next-generation energy reactor.

Recently, an excellent irradiation resistance has been reported in CoCrFeMnNi[10-13], which is so-called high-entropy alloys (HEAs). HEAs are typically defined as alloys containing at least five elements with concentrations between 5% and 35% [14]. HEAs have high configurational mixing entropy ($\Delta S_{mix}$), defined as $\Delta S_{mix} = -R\Sigma_i c_i \ln c_i$, where $c_i$ and $R$ are the compositional ratio and the gas constant, respectively[14,15]. HEAs have recently attracted much attention in the fields of materials science and engineering because they exhibit excellent performance under extreme conditions[15,16]. Even though the reason for its excellent irradiation resistance is still unclear, the atomic-level stress and local lattice distortions in the CoCrFeMnNi HEA have been believed to increase the migration barrier of point defects induced by irradiation[10]. These findings make the HEA material as a promising candidate in a high-energy particle irradiation environment.

Furthermore, the discovery of a HEA superconductor of Ti-Zr-Hf-Nb-Ta in 2014 by Koželj et al. triggered the development of HEA superconductors[17]. Since 2018, we have developed HEA-type superconducting compounds, in which the HEA concept was extended to complicated compounds having two or more crystallographic sites[18]. Comparing the effects of HEA states on superconductors with various crystal structural dimensionality, we found that the disordering effects introduced by the HEA-type site in BiS$_2$-based layered system[19] and tetragonal $Tr$Zr$_2$ ($Tr$: transition metals) quasi-two-dimensional system[20-22] does not suppress its original $T_c$ in pure phases. These findings suggest that the HEA effects in layered superconductors seem to work positively or at least less negatively. We also previously reported the synthesis of HEA-type *RE*123 polycrystalline samples and reported that the increase in $\Delta S_{mix}$ does not suppress superconducting properties including $J_c^{global}$ [23] and possible improvement of intra-grain $J_c$ ($J_c^{local}$) for *RE*123 using lighter *RE* elements including Dy, Ho [24]. These background knowledges have motivated us to study superconducting properties of epitaxially-grown high-entropy (HE) *RE*123 thin films. In this letter, we show the successful fabrication of the epitaxial thin film of *RE*123 on a SrTiO$_3$ substrate with various $\Delta S_{mix}$ at the *RE* site. To the best of our knowledge, this is the first report on fabrication of a HEA-type superconducting thin film. Although the $J_c$ of some films with more than three *RE* elements was slightly lower than that of pristine one, we found that all films exhibited the $J_c$ values over an order of 1.0 MA/cm$^2$ under conditions of $T <$ 20 K and $\mu_0 H < 7$ T. These results revealed a potential for the practical use of HE *RE*123 superconducting films, for example in the next-generation energy reactor.

All thin film samples were fabricated at 920°C by a pulsed laser deposition (PLD) method with a 266 nm Nd:YAG laser and frequency of 4 Hz with an oxygen pressure of $1.0 \times 10^1$ Pa during the growth, and then slowly cool down to room temperature at a rate of 5°C /min with the oxygen pressure of $2.0 \times 10^4$ Pa. Targets of *RE*Ba$_2$Cu$_3$O$_{7-\delta}$ (*RE*: Y, Sm, Eu, Dy, Ho) were prepared by the conventional solid state reaction in air as described in Refs. 23 and 24. The films were grown on a SrTiO$_3$ (STO) (001) substrate. X-ray diffraction (XRD) patterns were collected on a MiniFlex-600 diffractometer (RIGAKU), equipped with a D/tex-Ultra high-resolution detector, with a Cu-K$\alpha$ radiation by a conventional $\theta$-$2\theta$ method. The actual composition of the synthesized polycrystalline samples was investigated by energy dispersive X-ray spectroscopy (EDX) on a scanning electron microscope (SEM), TM-3030 (Hitachi), with Swift-ED (Oxford). The obtained compositions are listed in Table I. Thickness of the obtained films were estimated by Atomic Force Microscopy (AFM, Park System NX10) in a non-contact scanning mode. The superconducting properties were investigated using a superconducting quantum interference device (SQUID) magnetometer on MPMS-3 (Quantum Design). The temperature dependence of magnetization was measured after both zero-field cooling (ZFC) and field cooling (FC) with an applied field of 10 Oe.

According to the number of *RE* elements solved in the *RE* site, examined samples are labeled *RE*-1–*RE*-5 (see Table I). Figure 1a shows an X-ray $\theta$-$2\theta$ scan patterns of the obtained thin films for *RE*-1 to *RE*-5 grown on the STO (0 0 1) substrate. Except for the diffraction peak from the STO substrate, only (0 0 $l$) diffraction peaks of *RE*123 are observed, indicating the single phase and the $c$-axis oriented growth of obtained thin films. Slight broadening of the (0 0 $l$) diffraction peaks was observed (Fig. 1b) for films with a larger number of *RE* elements, while no peak split was seen by laboratory XRD.

Through EDX and SEM by back scattered electron scanning mode analysis (Fig. 2), we confirmed that the actual composition of the films is comparable to the nominal value and there is no compositional segregation. The estimated compositions are summarized together with $\Delta S_{mix}$ value in Table I. Although the fabricated films have $\Delta S_{mix}$ lower than 1.5$R$, which is a typical value of HEAs, the highest $\Delta S_{mix}$ among the films is 1.42$R$, which is close to that of HEAs and enough to examine the effect of entropy to superconducting properties of the *RE*123 films.

Figure 3(a) shows the temperature dependence of magnetization for *RE*-1–*RE*-5. Large diamagnetic signals corresponding to the superconductivity are observed at around 87 K for all samples. Slightly lower $T_c$ of obtained samples as compared to those of previously reported bulk samples[24] would be related to the oxygen amount condition during growth and/or cooling to room temperature. The bulk nature of superconductivity of obtained samples were confirmed through the *M-T* measurements and magnetization-magnetic field (*M-H*) loops in Fig. 4. Although a HE thin film exhibited same $T_c$ as pristine one, broadening of transition below $T_c^{mag}$ was observed in *RE*3–*RE*5 samples. The slight broadening of transition is also observed in some HEA-type compound superconductors[22,25]. From the specific heat measurement of those superconductors, broadening of superconducting transition jump was found, indicating that the observed broadening of magnetization transition would be related to similar origins. To reveal the origin of these phenomena, further experiments which can directly observe the superconducting gap structure, such as scanning tunneling microscopy (STM) or Angle-resolved photoemission spectroscopy (ARPES) measurement are desired. In that sense, the successful fabrication of thin film is a meaningful achievement for such investigations.

To estimate $J_c$ of the thin films, *M-H* loops were measured. From the obtained *M-H* loops, $J_c$ was estimated using the Bean's model[26]: $J_c = 20\Delta M / b(1-b/3a)$ (A/cm$^2$), where $a$ and $b$ are lengths determined by sample shape, and $\Delta M$ is obtained from the width of the $M$-$H$ curve. The typical results on $J_c$ at $T$ = 4.2 K are plotted as a function of magnetic field in Fig. 3(b). At lower magnetic fields, the difference of $J_c$ between the samples with zero to high entropy are clear, while the difference becomes smaller in higher magnetic fields except for $RE$-3. In particular, the $J_c$ of $RE$-4 and $RE$-5 closed to that of $RE$-1 above $\mu_0 H$ = 2.0 T. As shown in Fig. S1 (See the Supplemental material), $J_c$ against number of $RE$ at low-magnetic-field region shows a similar trend with previously reported $J_c^{local}$ in polycrystalline samples[24]. $M$-$H$ loop and $J_c$ at $T$ = 2.0, 4.2, 10.0, 20.0, 50.0, 77.3, 90.0, 100.0 K for $RE$-1–$RE$-5 are plotted in Fig. 4(a-h). Although $J_c$ of $RE$-3–$RE$-5 slightly decreases compared to that of $RE$-1 at low magnetic fields, the $J_c$ values of the $RE$-4 and $RE$-5 films recorded approximately 2.4 and 2.3 MA/cm$^2$ at $T$ = 4.2 K under around $\mu_0 H$ = 7 T. In addition, their $J_c$ also exceed an order of 1.0 MA/cm$^2$ up to 20 K in entire fields. We emphasize that the finding that the increase of $\Delta S_{mix}$ does not significantly deteriorate the $J_c$ should be quite positive because HE effects are expected to improve functionality other than superconducting properties. As described earlier, the improvement of irradiation resistance is one of the most important tasks for the practical use of high-$T_c$ superconductors in a fusion reactor. Therefore, our current results will encourage further studies on HE $RE$123 materials from the perspective of possible achievement of both high-$J_c$ and excellent irradiation tolerance. Moreover, it is well known that an enhancement of $J_c$ is usually achieved by combining with introduction of artificial pinning center (APC) such as BaZrO$_3$[27-29]. Since the present samples are not optimized with any APCs, there should be more space for further improvement of $J_c$ in HE $RE$123 films. Further investigation on film fabrication process, superconducting properties, and irradiation tolerance is required to clarify the irradiation resistance of HE $RE$123 superconductors and for further development of superconducting application field.

In conclusion, we reported the successful fabrication of high-entropy $RE$Ba$_2$Cu$_3$O$_{7-\delta}$ ($RE$123, $RE$: rare earth) epitaxial thin films on the SrTiO$_3$ substrate by pulsed laser deposition method. Observation of only (0 0 $l$) diffraction peaks of $RE$123 except for substrate indicates the single phase and the $c$-axis oriented growth of obtained thin films. SEM-EDX analyses revealed that there is no compositional segregation in all films and confirmed different configurational entropy of mixing at the $RE$ site. From the characterizations of superconducting properties, the $J_c$ of the $RE$-4 and $RE$-5 films are estimated as approximately 2.4 and 2.3 MA/cm$^2$ at 4.2 K under around 7 T and also exceeded an order of 1.0 MA/cm$^2$ up to 20 K in entire fields. In addition, their $J_c$ becomes comparable to that of $RE$-1 under high-field region. Since the increase of $\Delta S_{mix}$ does not deteriorate the $J_c$ significantly in HE $RE$123 films, this material will be useful for application under high fields and irradiation like a next-generation fusion reactor, after improvement of irradiation tolerance.

**Acknowledgements**
The authors thank T. Nakano, O. Miura, Y. Shukunami, and Y. Goto for supports in experiments and discussion. This work was partly supported by JSPS-KAKENHI (18KK0076) and Tokyo Metropolitan Government Advanced Research (H31-1).

**Authors' contributions**
A.Y. and Y.M. designed the research. A.Y. and T.M. carried out PLD fabrication of thin films. A.Y. prepared target samples and characterized obtained films. A.Y. carried out $M$-$T$ and $M$-$H$ measurements. A.Y. and K.H. and Y.N. carried out AFM measurement. A.Y. and Y.M. analyzed magnetization data. A.Y. and Y.M. wrote the manuscript. All the authors made contributions to writing the manuscript.

**Conflict of interest**
The authors declare that they have no conflict of interest.

Table I. Compositional, configurational mixing entropy ($\Delta S_{mix}$), and superconductivity, thickness of $RE$Ba$_2$Cu$_3$O$_{7-\delta}$ thin films

| $RE$ | $RE$ composition | $\Delta S_{mix}/R$ | $T_c$ (K) | $J_c$ (MA/cm$^2$) $T$ = 4.2 K, $H$ = 7.0 T | $t$ (nm) |
|---|---|---|---|---|---|
| 1 | Y | 0 | 86.0 | 2.5 | 220 |
| 3 | Y$_{0.49}$Sm$_{0.27}$Eu$_{0.23}$ | 1.04 | 87.6 | 1.5 | 255 |
| 4 | Y$_{0.35}$Sm$_{0.18}$Eu$_{0.21}$Dy$_{0.26}$ | 1.35 | 87.8 | 2.4 | 181 |
| 5 | Y$_{0.25}$Sm$_{0.10}$Eu$_{0.10}$Dy$_{0.10}$Ho$_{0.45}$ | 1.42 | 87.0 | 2.3 | 208 |

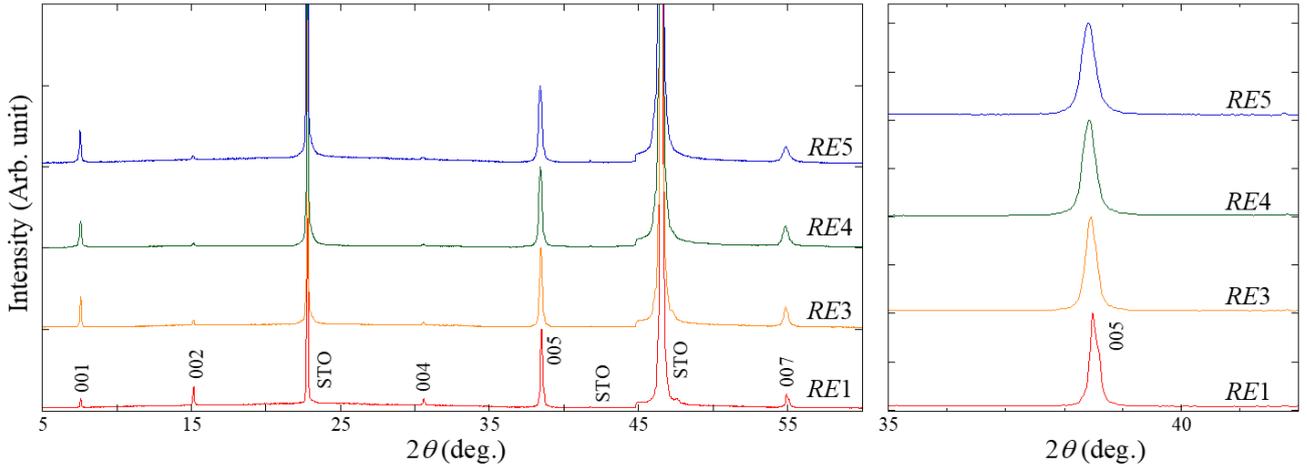

FIG. 1. X-ray $\theta$–2$\theta$ scan patterns of the obtained $RE$Ba$_2$Cu$_3$O$_{7-\delta}$ thin films for $RE$1 to $RE$5 grown on STO$_3$ (0 0 1).

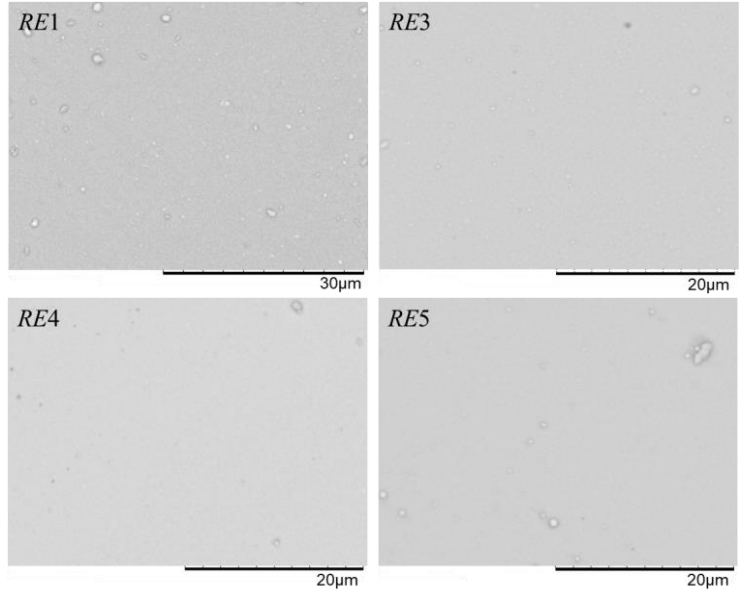

FIG. 2. Back scattering electron (BSE) images of the obtained $RE$Ba$_2$Cu$_3$O$_{7-\delta}$ thin films for $RE$1 to $RE$5 grown on STO$_3$ (0 0 1)

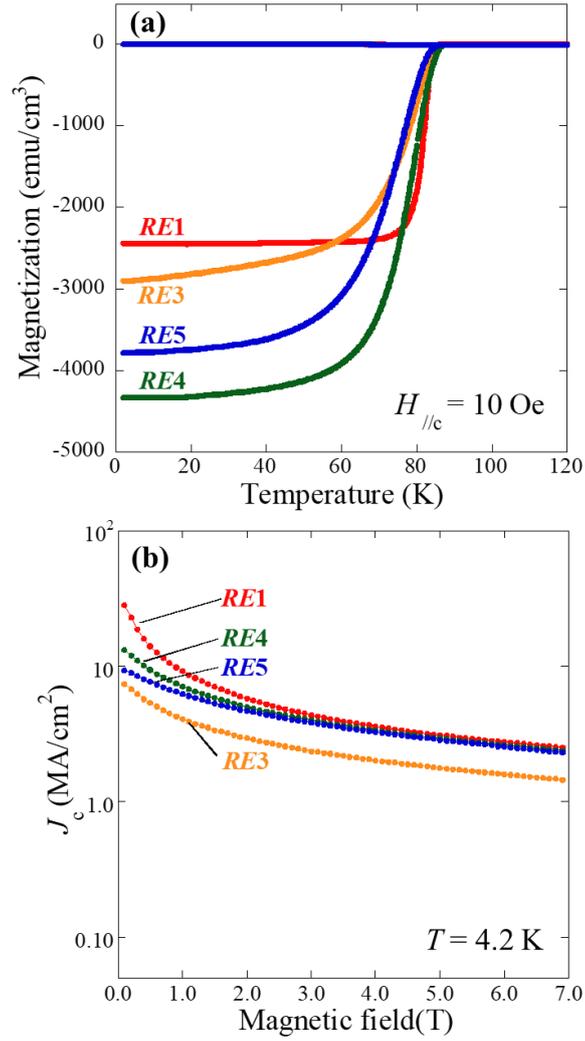

FIG. 3. (a) Temperature dependence of magnetization for the obtained $RE$Ba$_2$Cu$_3$O$_{7-\delta}$ thin films for $RE$1 to $RE$5. (b) Magnetic field dependences of magnetic $J_c$ at $T$ = 4.2 K estimated using Bean's model.

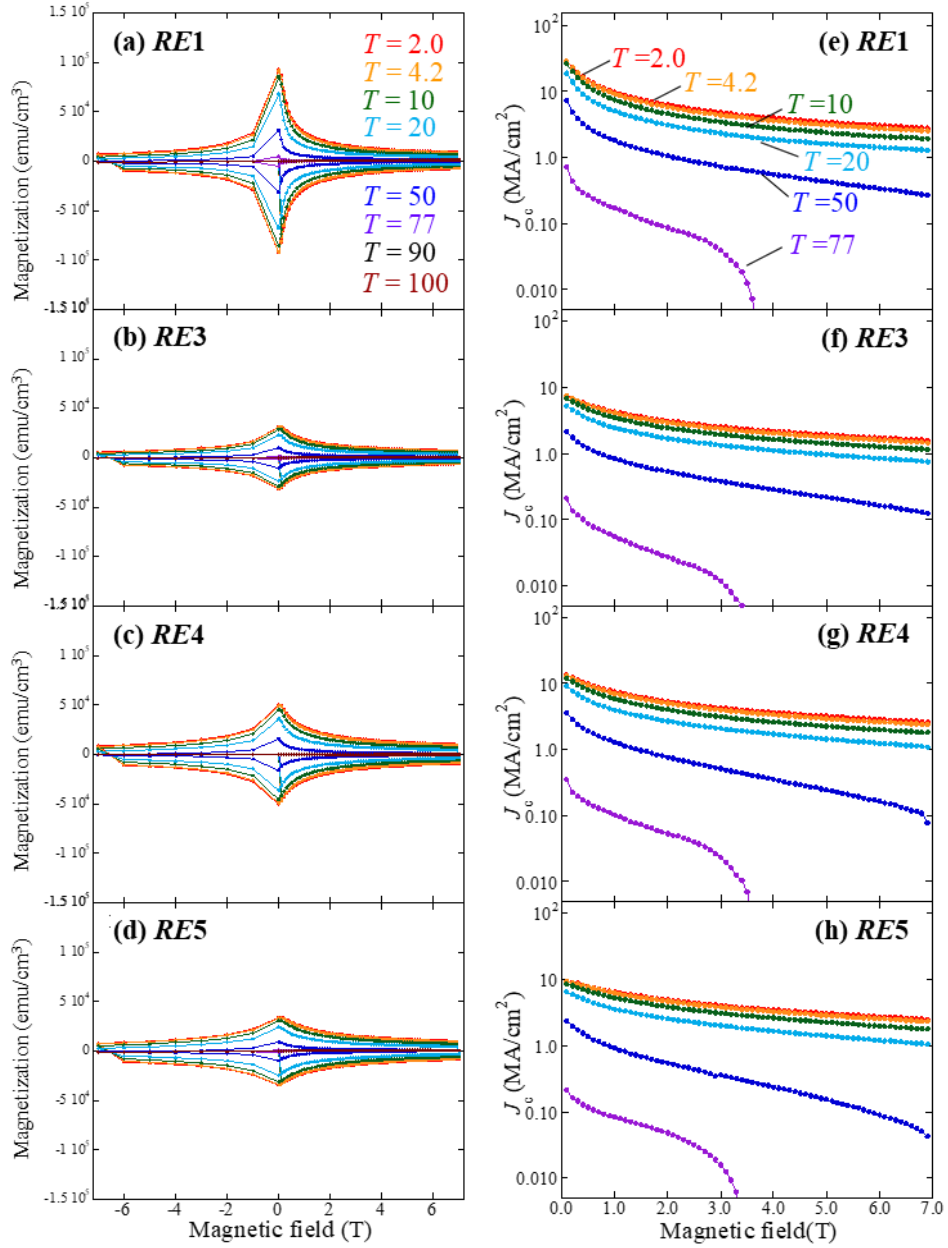

**FIG. 4. (a-d)** Magnetic field dependences of magnetization at $T = 2.0-100$ K for the obtained $RE$Ba$_2$Cu$_3$O$_{7-\delta}$ thin films for $RE$1 to $RE$5. **(e-h)** Magnetic field dependences of magnetic $J_c$ at $T = 2.0-77$ K estimated using Bean's model.

7